\documentclass[preprint,showpacs,showkeys,preprintnumbers,amsmath,amssymb]{revtex4}

\usepackage[pdftex]{graphicx}
\usepackage{float}
\usepackage{xcolor}
\usepackage{subcaption}
\usepackage{dcolumn}
\usepackage{bm}
\usepackage{multirow}
\usepackage{array}
\usepackage{siunitx} 
\usepackage{booktabs}
\usepackage{calligra}
\DeclareMathAlphabet{\mathcalligra}{T1}{calligra}{m}{n}
\DeclareFontShape{T1}{calligra}{m}{n}{<->s*[2.2]callig15}{}
\newcommand{\ra}[1]{\renewcommand{\arraystretch}{#1}}

\newcolumntype{P}[1]{>{\centering\arraybackslash}p{#1}}

\AtBeginDocument{
\heavyrulewidth=.08em
\lightrulewidth=.05em
\cmidrulewidth=.03em
\belowrulesep=.65ex
\belowbottomsep=0pt
\aboverulesep=.4ex
\abovetopsep=0pt
\cmidrulesep=\doublerulesep
\cmidrulekern=.5em
\defaultaddspace=.5em
}

\begin{document}
\bibliographystyle{ieeetran}

\preprint{APS/123-QED}

\title{Opinion Dynamics in Financial Markets via Random Networks}

\author{Mateus F. B. Granha}
 \email{mateus.granha@upe.br}
\affiliation{F\'isica de Materiais, Universidade de Pernambuco, Recife, PE 50720-001,
  Brazil}
 
\author{Andr\'e L. M. Vilela}
 \email{andre.vilela@upe.br}
\affiliation{Universidade de Pernambuco, Recife, PE 50720-001,
  Brazil}
\affiliation{Center for Polymer Studies and Department of Physics,
  Boston University, Boston, MA 02215, USA}
  
\author{Chao Wang}
 \email{chaowanghn@vip.163.com}
\affiliation{College of Economics and Management, Beijing University of Technology, Beijing, 100124, China}

\author{Kenric P. Nelson}
\email{kenricpn@bu.edu}
\affiliation{Photrek LLC, Watertown, MA 02472, USA}%

\author{H. Eugene Stanley}%
 \email{hes@bu.edu}
\affiliation{Center for Polymer Studies and Department of Physics,
  Boston University, Boston, MA 02215, USA} 

\date{\today}

\begin{abstract}

\noindent
We investigate the financial market dynamics by introducing a heterogeneous agent-based opinion formation model. In this work, we organize the individuals in a financial market by their trading strategy, namely noise traders and fundamentalists. The opinion of a local majority compels the market exchanging behavior of noise traders, whereas the global behavior of the market influences the fundamentalist agents' decisions. We introduce a noise parameter $q$ to represent a level of anxiety and perceived uncertainty regarding the market behavior, enabling the possibility for an adrift financial action. We place the individuals as nodes in an Erdös-Rényi random graph, where the links represent their social interaction. At a given time, they assume one of two possible opinion states $\pm 1$ regarding buying or selling an asset. The model exhibits such fundamental qualitative and quantitative real-world market features as the distribution of logarithmic returns with fat-tails, clustered volatility, and long-term correlation of returns. We use Student's t distributions to fit the histograms of logarithmic returns, showing the gradual shift from a leptokurtic to a mesokurtic regime, depending on the fraction of fundamentalist agents. We also compare our results with the distribution of logarithmic returns of several real-world financial indices.
\end{abstract}



\maketitle

\section{INTRODUCTION} \label{intro}

For decades now, especially over the last years, financial markets and economic systems have fascinated researchers and investors worldwide. The effectiveness of methods and techniques from Statistical Mechanics has been a compelling element for the comprehension of financial dynamics as a complex system \cite{stanley2000introduction, Stanley2002, Farmer2005, Baldovin2007, Li2020}. Simultaneously, it has been responsible for developing the interdisciplinary field of Econophysics, where agent-based models have a significant contribution. In this framework, several models use a set of elementary rules to represent the agent behavior and interactions. Thus, yielding intense emergent collective phenomena \cite{bonabeau2002agent, macal2005tutorial, macal2009agent, vilela2019majority, zubillaga2019three, Feng2012, Zhao2011}.

Oliveira \cite{de1992isotropic, Oliveira1993} pioneered in the study of opinion dynamics modeled by the majority-vote model (MVM) with noise, a sociological adaptation of a magnetic spin model that presents similar critical behavior. Lima et al., Campos et al. and Pereira et al. \cite{lima2008majority, campos2003small, pereira2005majority} investigated the effects of complex networks on the MVM social dynamics, thus revealing an ordered phase over the increasing complexity of these networks. Vilela et al. \cite{vilela2019majority} incorporated the rational and emotional behavior of agents in financial markets introducing the global-vote model, inspired by the majority-vote dynamics. 

The MVM consists of a comprehensive approach to the study of social dynamics with interactions embedded in regular and complex networks \cite{de1992isotropic, Oliveira1993, santos1995anisotropic, vieira2016phase, pereira2005majority, lima2008majority, campos2003small, vilela2009, vilela2018effect, vilela2021}. In this agent-based model with two states, the individual's opinion in a given time instant may assume one of the two values, $\pm 1$, regarding some social discussion. Furthermore, an agent in the social network assumes the opinion of the majority of its neighboring spins with probability $(1 - q)$ and the opposite opinion with probability $q$. The variable $q$ stands for the noise parameter of the model and measures the social unrest or social temperature of the system. The MVM exhibits a second-order phase transition for a critical noise value $q = q_c \approx 0.075$ in square lattice networks of social interactions \cite{de1992isotropic}.

Complex Networks consist of a natural preference to the study of real-world complex systems, such as climate analysis, biological neural connections, World Wide Web, public transportation, airline networks, financial markets, among others \cite{feldhoff2015complex, reijneveld2007application, barabasi2000scale, an2014synchronization}. In this context, the random graph networks describe the topology of a set of $N$ nodes connected with chance $p$, thus introducing a fundamental probability distribution over graphs. The Erdös-Rényi algorithm is one well-known method for the assembly of random graph networks. The method connects an initial set of $N$ isolated nodes by adding a total of $pN(N-1)/2$ random links between them, while double connections are forbidden \cite{erdos1960evolution, bollobas1985random}.

\begin{figure}[H]
	\centering
	\includegraphics[scale=0.22]{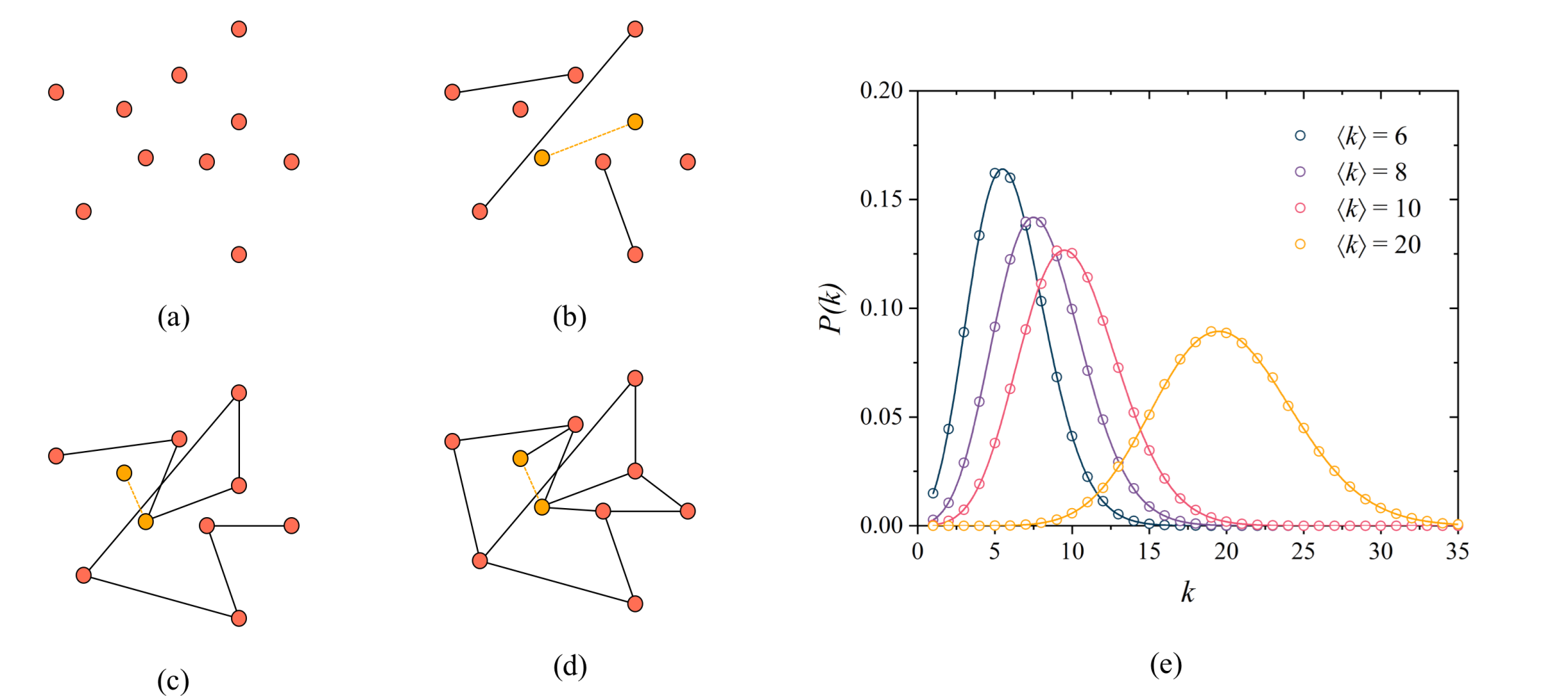}
	 \caption{Illustration of the Erdös-Rényi process of building random graph networks. From (a) to (d), $N = 10$ isolated nodes are connected by adding a total of $pN(N-1)/2$ random links between them. While double connections are forbidden, the final network has $\left<k\right> = 3$. In (e) we show the degree distribution $P(k)$ averaged over ten networks for $\left<k\right> = 6, 8, 10$ and $20$ with $N = 2 \times 10^4$. The lines represent Poisson fits for the data.}
	 \label{RG representation}
\end{figure}

In Figure \ref{RG representation}, we illustrate the Erdös-Rényi process of building a random graph network with $N = 10$ and $\left<k\right> = 3$. We also present the averaged degree distribution $P(k)$ over ten networks with $N = 2 \times 10^4$ nodes and several values of $\left<k\right>$. The lines correspond to Poisson fits for the data with average degree $\left<k\right> = pN$, in agreement for the degree distribution of random graphs for large networks.

\vspace{0.8cm}

Modern financial markets operate as a complex system. Its dynamics depend not only on the rational strategy of the individual but also on the emotional circumstances that define the investor's psyche. At first, individual decisions might seem challenging to model; still, social agents tend to follow a herding behavior, as they feel sheltered when the crowd is endorsing their choices. In this way, trader social dynamics is reasonably attainable \cite{cont2000herd, sznajd2002simple}. This bias to follow a significant group is often embraced by less experienced agents, denominated \textit{noise traders}. They are highly susceptible to the dominant opinion and tend to overreact to fresh news regarding buying or selling. 

In contrast, other traders, denominated noise contrarian traders or fundamentalists, follow the global minority in a given market as an investment strategy. Thus they buy when stock prices decline and sell when the prices increase \cite{sznajd2002simple, lux1999scaling, bornholdt2001expectation, kaizoji2002dynamics, takaishi2005simulations}. Therefore, their decision-making process drives the asset price to its fundamental value.  

Inspired by the global-vote model for financial markets \cite{vilela2019majority}, we propose to investigate the influence of random graph networks on the time evolution of financial markets quantity measures. This work employs an opinion formation model to study the economic mechanics of noise traders and contrarian individuals interacting through a random graph of social influence.

This work is organized as follows. In Section \ref{secmodel}, we describe the global-vote opinion formation model for financial markets and present the relevant quantities analyzed in our simulations. In Section \ref{results}, we present the numerical results obtained along with the corresponding discussions. Section \ref{conclusion} concludes with our final remarks.

\section{THE MODEL} \label{secmodel}

We represent the financial agents on the market and their interactions using a complex network structure. We place $N$ agents on the nodes of a random graph network, where the links represent the interaction between the agents in the market.  We map the agent's financial decision (opinion or option) at a given time $t$ by a spin variable, which may assume one of two values: $+1$ or $-1$, regarding buying or selling an asset. Furthermore, to model the essential dynamics of real-world financial markets, we randomly distribute two sets of individuals: a fraction $f$ of fundamentalist traders, also referred to as noise contrarians, and the remaining fraction $1 - f$ of noise traders. We use the spin variables $\lambda$ and $\alpha$ to stand for the financial option of noise traders and contrarian agents, respectively.

We introduce a noise parameter variable $q$ to model a financial level of anxiety and perceived uncertainty present in the market. In this study, $q$ influences contrarian and noise traders' decisions and represents the probability of performing an inaccurate financial action. In other words, $q$ is the chance for an agent to choose the opposite of his standard strategy when negotiating in the financial market.

\subsection{Noise Traders} \label{ssecmodel1}

A noise trader individual $i$, with opinion $\lambda_i$, assumes the same opinion of the majority of its neighbors with probability $1 - q$, and the opposite option with probability $q$. We write the opinion flipping probability for the noise trader agent $i$ as follows

\begin{equation}
	\omega (\lambda_{i}) = \frac{1}{2} \left[1 - (1 - 2q) c_\lambda \lambda_{i} \textrm{sgn} (m)\right],
	\label{wlambda}
\end{equation}
where $\textrm{sgn}(x) = -1, 0, +1$ for $x < 0$, $x = 0$ and $x > 0$, respectively. Here, $c_\lambda$ stands for the agent strategy, where we set $c_\lambda = +1$ to model the noise traders trend to agree with the local majority of their neighbors. The variable $m$ quantifies the local predominant opinion, or local magnetization, defined as

\begin{equation}
m = \sum_{\delta = 1}^{k_{i}} \lambda_{i + \delta}.
\end{equation}
The summation is over all the $k_{i}$ neighboring agents connected to the trader at node $i$, with opinion $\lambda_{i}$. From Eq. (\ref{wlambda}), when $q = 0$ the noise trader adopts its local predominant opinion. When we increase the market anxiety parameter $q$, the noise trader tends to follow the opposite opinion of its local majority.

\subsection{Noise Contrarians} \label{ssecmodel}

A fundamentalist agent tends to follow the market's minority opinion with probability $1 - q$, while following the majority opinion with chance $q$. We define the prevailing option of the system as a global magnetization, which influences noise contrarian traders' financial option. The option of a noise contrarian agent $j$, $\alpha_j$, flips with probability

\begin{equation}
	\omega (\alpha_{j}) = \frac{1}{2} \left[1 - (1 - 2q) c_\alpha \alpha_{j} \textrm{sgn}(M) \right],
	\label{walpha}
\end{equation}
where $c_\alpha$ is defined as the fundamentalist's strategy, and since these individuals tend to agree with the global minority of the system, $c_\alpha = -1$, for all noise contrarian agents.

The variable $M$ measures the average market opinion of the system, thus revealing the economic order. The magnetization $M$ accounts for the financial opinion of every agent on the market, and is evaluated as

\begin{equation}
	M = \frac{1}{(N_\lambda + N_\alpha)} \left(\sum_{i=1}^{N_\lambda} \lambda_{i} + \sum_{j=1}^{N_\alpha} \alpha_{j}\right),
	\label{systemmagnet}
\end{equation}
where $N_\lambda = N(1 - f)$ stands for the number of noise trader agents in the network and $N_\alpha = Nf$ represents the number of noise contrarian agents. For $M = 1$, every agent on the market has an option equal to $+1$. Similarly, $M = -1$ denotes a market configuration where every agent opinion is equal to $-1$. $M = 0$ represents the case where half of the agents have opinion $+1$ and the other half $-1$. Additionally, intermediate values of $M$ infer the dominant opinion of the market.

We remark that when $q = 0$, a contrarian agent always follow the opposite opinion of the global magnetization: buy (sell) when the majority sells (buy). As $q$ increases, there is a greater probability for the contrarian agent to follow the majority, adopting the opposite of his inherent financial market strategy.

In the global-vote dynamics for financial markets, we recover the standard majority-vote model on random graphs when the fraction of contrarians $f$ is zero. In this case, we observe a order-disorder phase transition at a critical point $q_c$ for each value of the average connectivity $\left<k\right>$ in the thermodynamic limit $N \rightarrow \infty$ \cite{pereira2005majority}. This $f = 0$ system presents an ordered phase, with large clusters of agents that share the same opinion for values of noise $q$ below the critical point $q_c$. By increasing $q$, the same opinion clusters fade, and the magnetization (order parameter) approaches zero. 

In this work, we analyze the global-vote dynamics for financial markets on random graphs for several values of the average connectivity $\left<k\right>$ and noise contrarian traders fraction $f \neq 0$. We set $q$ near $q_c(f = 0, \left<k\right>)$ to model real-world market dynamics adequately since previous investigations suggest the strong market phase emerge when the system is close to its critical melting point \cite{bornholdt2001expectation, vilela2019majority}.

\section{RESULTS AND DISCUSSION} \label{results}

We perform Monte Carlo simulations on Erdös-Rényi random graphs of size $N = 10201$ and several values of the average connectivity $\left<k\right>$. For the network of financial interactions, we randomly place a fraction $1-f$ of noise trader on the network nodes, and we occupy the remaining fraction $f$ with noise contrarian agents. In this way, $N = N_{\lambda} + N_{\alpha}$.

This work considers that a real-world market presents a small fraction of noise contrarian agents. To this extent, we investigate the influence of the noise contrarian agents for $q$ near its critical value for the $f = 0$ case, obtained in previous studies within several values of $\left<k\right>$ \cite{pereira2005majority}, where our model exhibits key real-world market features \cite{vilela2019majority}. Additionally, we consider small values for the average number of connections of an agent in a market, i.e., small values of $\left<k\right>$, supported by the agreement between the data of real-world markets and the results of our simulations.

The market dynamics evolve as follows. We select a randomly chosen agent and update its opinion accordingly with the probabilities given by Eqs. (\ref{wlambda}) or (\ref{walpha}) for a noise trader or a noise contrarian agent, respectively. We repeat this process $N$ times, thus defining a unity of time in one Monte Carlo step (MCS). This way, each agent's opinion is updated once on average on one MCS. To discard the transient regime, we allow this dynamic to run during $10^3$ MCS, and we perform our analysis in the subsequent $10^5$ MCS.

In the financial context, we shall relate the global magnetization of the system to the aggregate excess demand of a particular asset, for its impact on stock prices \cite{bornholdt2001expectation, kaizoji2002dynamics, takaishi2005simulations}. Thus a positive demand ($M > 0$) causes prices to rise, while a negative demand ($M < 0$) causes prices to fall. Moreover, markets that fluctuate around equilibrium \cite{cont2000herd} exhibit an average excess demand that oscillates around zero.

The financial return time series represents the price variation of a given asset over time. Positive (negative) returns relate to profit (loss) during the period of analysis. We quantify the logarithmic return at a given time $t$ as follows

\begin{equation}
	r(t) = \textrm{log}\left[\left| M(t) \right|\right] - \textrm{log}\left[\left| M(t-1) \right| \right].
\end{equation}

\begin{figure}[H]
	\centering
	\includegraphics[scale=0.35]{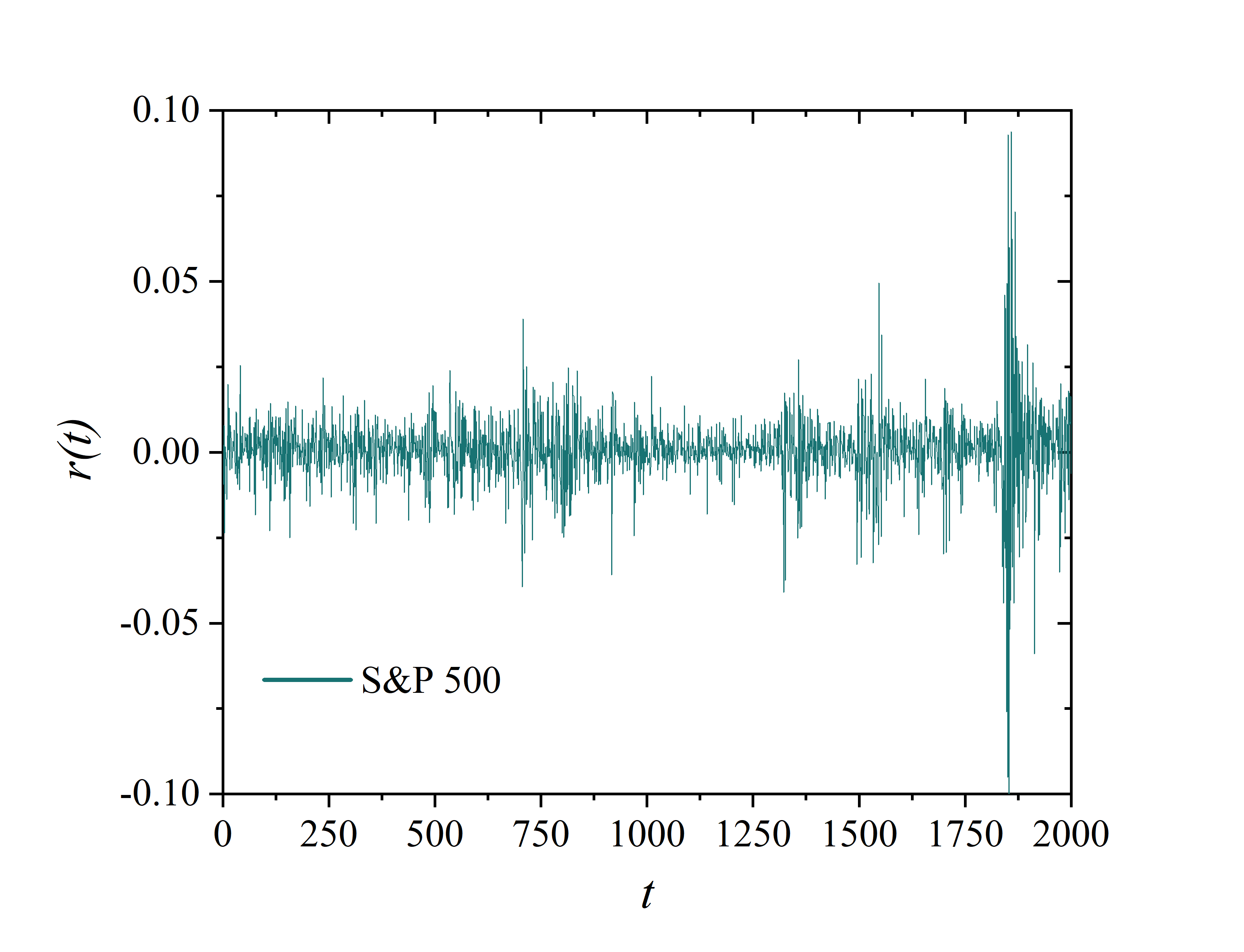}
	 \caption{Logarithmic returns for the daily closing price of S\&P 500 index in US dollars from Nov 01, 2012 to Dec 08, 2020. The high volatility observed for the period around t = 1850 represents the Coronavirus Stock Market Crash of 2020.}
	 \label{fig: SP500 Ret}
\end{figure}

In Figure \ref{fig: SP500 Ret}, we present the logarithmic return for the closing values of the S\&P 500 index in US dollars, from Nov 01, 2012 to Dec 08, 2020. Similar to any other traditional asset, the S\&P 500 index depends mainly on market supply and demand, a fundamental financial mechanism that yields several periods of strong return variations. Periods of significant fluctuations of returns are compressed in time, denoting the clustered volatility effect for the analyzed index price. This financial phenomenon is known as volatility clustering, and it can be elucidated by Mandelbrot's observation that ``large changes tend to be followed by large changes - of either sign - and small changes tend to be followed by small changes'' \cite{mandelbrot1963variation}.

Figure \ref{fig: SP500 Ret} also shows that the period around t = 1850 presents the most considerable time-series return volatility. It denotes the Coronavirus pandemic phase, where government officials halted the economic activity. The panic and uncertainty triggered by the financial impacts of such measures led to a expressive stock market crash \cite{vega2021, wei2021}.

\begin{figure}[ht]
  \centering
	\includegraphics[scale=0.07]{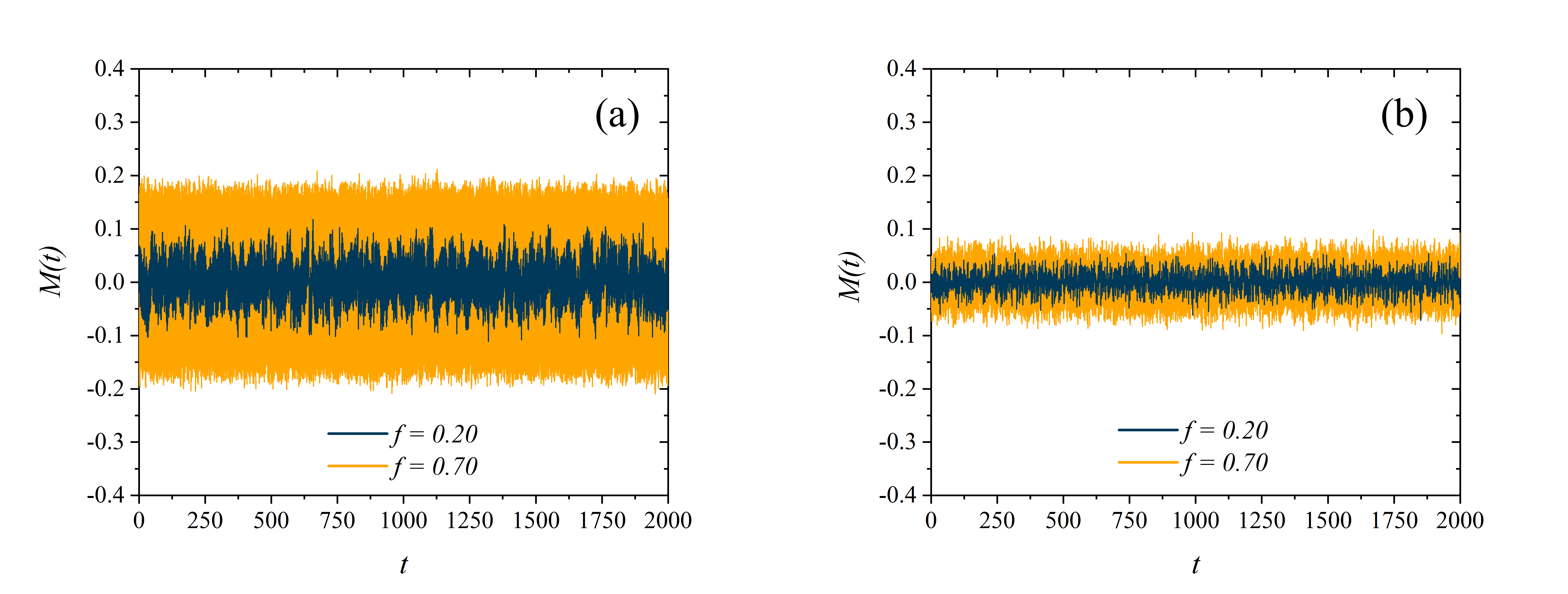}
  \caption{Time series of the order parameter $M$ for two sets of parameters: (a) $\left<k\right> = 6$ for $q = 0.240$, and (b) $\left<k\right> = 50$ for $q = 0.411$.}
  \label{Magnetizations}
\end{figure}

In Figure \ref{Magnetizations}, we analyze the influence of the average connectivity $\left<k\right>$ and the noise contrarian fraction $f$ on the market order of the model on random graphs. Figure \ref{Magnetizations}(a) exhibits two distinct market phases: a ``strong market phase" for $f = 0.20$ (dark blue), where the system is highly volatile and the magnetization exhibits a irregular wave pattern; and a ``weak market phase" for $f = 0.70$ (yellow), where magnetization values are roughly randomly distributed \cite{vilela2019majority}. This result demonstrates that the increase of contrarians tends to stabilize the market dynamics, where the high volatile periods are present but to a substantially lower extent. Additionally, the increase of the agent's average number of connections $\left<k\right>$ also yields stochastic patterns for the market demand, but driving a contraction on the amplitude of the fluctuations, as observed in Figure \ref{Magnetizations}(b).

\begin{figure}[ht]
  \centering
\includegraphics[scale=0.07]{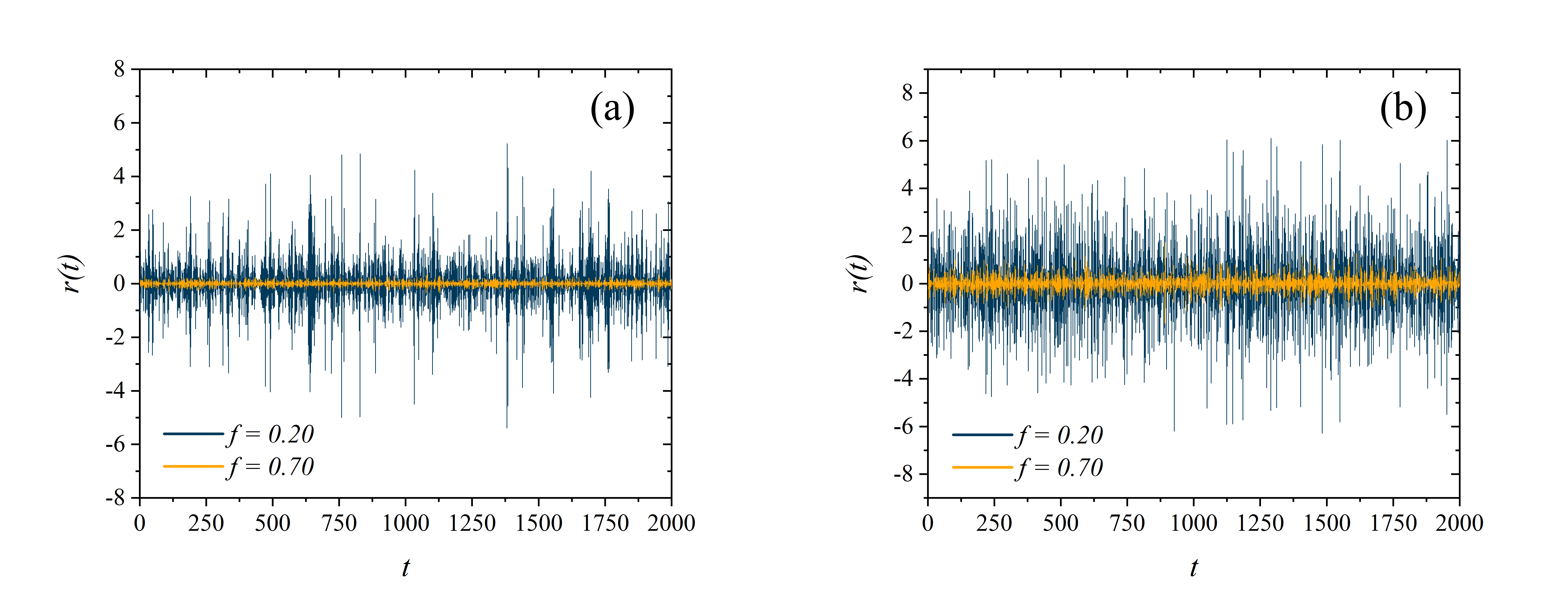}
  \caption{Logarithmic returns of the absolute magnetization for (a) $\left<k\right> = 6$ and $q = 0.240$, (b) $\left<k\right> = 50$ and $q = 0.411$.}
  \label{Return}
 \end{figure}

Figure \ref{Return}, displays the logarithmic returns of the absolute value of the order parameter. In financial markets, we define volatility as a measure of the variation around the average returns observed in an asset's time series. In this figure, we note intensive market fluctuations for $f = 0.20$, embodied by the large spikes in the plot. The result also shows clustered volatility, a real-world market feature. We observe that the presence of contrarians tends to stabilize the market, indicated by attenuated fluctuations of the returns. Furthermore, when comparing Figures \ref{Return}(a) and \ref{Return}(b), it becomes clear that increasing the average connectivity of the network will similarly increase the market's volatility and simultaneously deviate the behavior from the expected of real-world financial markets. Figure \ref{Return} also shows that periods of high volatility tend to be clustered, for lower values of $f$.

\begin{figure}[htpb]
  \centering
  \includegraphics[scale=0.07]{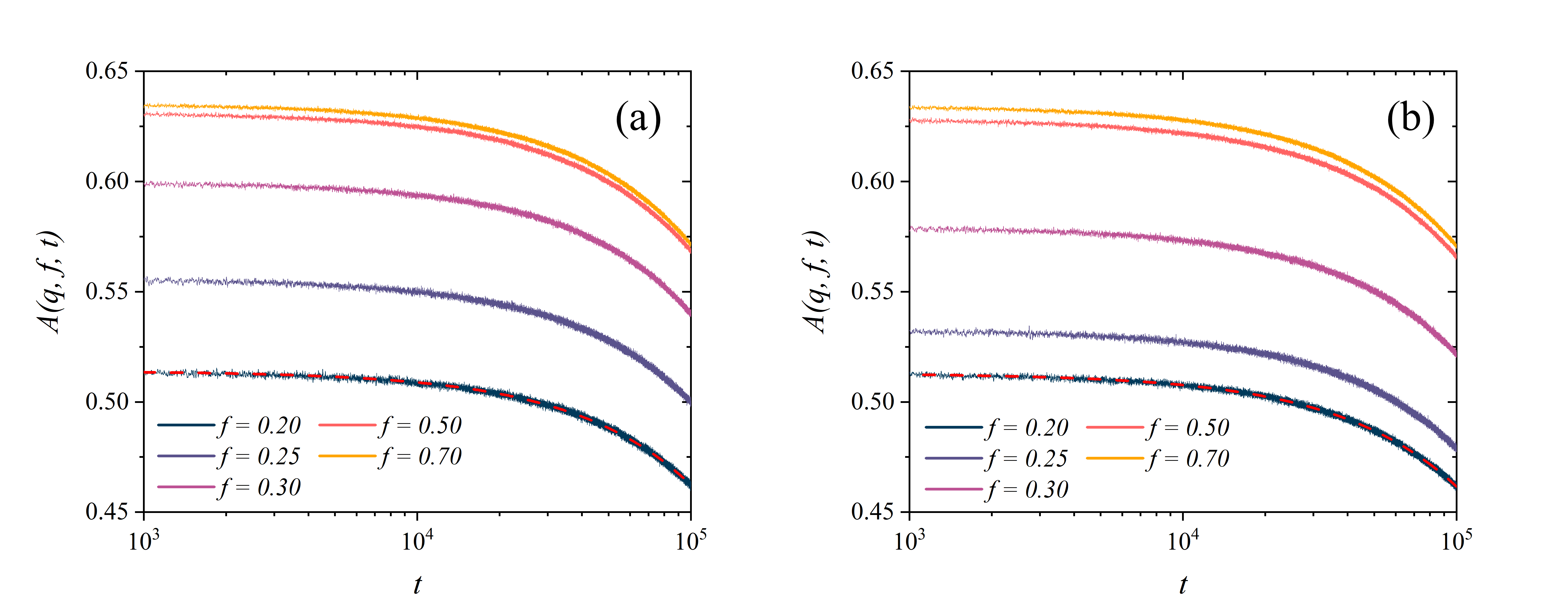}
  \caption{Linear-log plot of the autocorrelation of absolute logarithmic returns for (a) $\left<k\right> = 6$ and $q = 0.240$, and (b) $\left<k\right> = 8$ and $q = 0.275$. The dashed red lines represent exponential fits for the data.}
  \label{ACR}
\end{figure}

In order to quantify the effects of volatility clustering, we shall compute the autocorrelation of absolute returns as follows

\begin{equation}
	A(\tau) = \frac{\sum_{t = \tau +1}^{T} \left[\left|r(t)\right| - |\overset{-}{r}|\right]  \left[\left|r(t - \tau)\right| - |\overset{-}{r}|\right]}{\sum_{t = 1}^{T} \left[\left|r(t)\right| - |\overset{-}{r}|\right]^2},
	\label{Eq-ACR}
\end{equation}
where $1 \leq \tau \leq 10^5 \ \textrm{MCS}$ is the time-step difference between observations, $T = 10^5 \  \textrm{MCS}$ is the time of simulation, $r(t)$ is the return at a time $t$ and $\overset{-}{r}$ the average value of the return. The function defined by Eq. (\ref{Eq-ACR}) measures nonlinear correlations in observations of the absolute value of log-returns as a function of the time delay between them. Figure \ref{ACR} displays the autocorrelation for several values of $f$, and it shows that the returns present a strong correlation in time with exponential decay \cite{kaizoji2002dynamics, takaishi2005simulations, bornholdt2001expectation, vilela2019majority}.
 
\begin{figure}[htpb]
	\centering
	\includegraphics[scale=0.35]{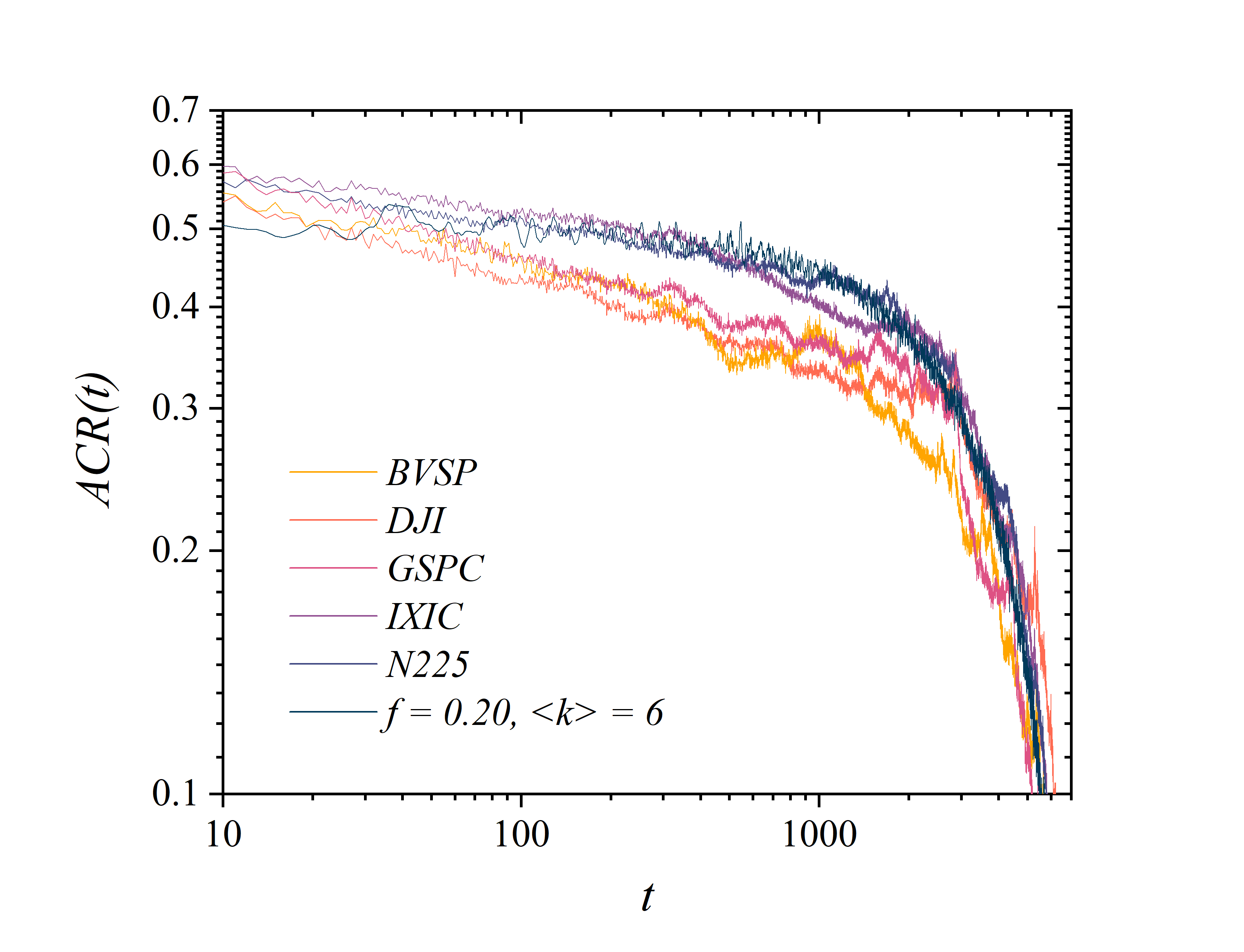}
	 \caption{Linear-log plot of the autocorrelation of absolute logarithmic returns of the closing values for several financial indices. Also shown is the autocorrelation of absolute log-returns for $\left<k\right> = 6$, $q = 0.240$ and $f = 0.20$ (dark blue).}
	 \label{ACR-RealWorld}
\end{figure}

To illustrate this exponential behavior, we perform an exponential fit of the autocorrelation function $A(q, f, t) \sim \textrm{exp}(-t/t_0)$ for $f = 0.20$, and we obtain $1/t_0 \approx 2.5 \times 10^{-7}$ and $1/t_0 \approx 3.0 \times 10^{-7}$ for Figures \ref{ACR}(a) and \ref{ACR}(b), respectively. Other values for the fraction of noise contrarians yield similar results.

In Figure \ref{ACR-RealWorld}, we compare the results of our model with real-world financial indices. We calculate the autocorrelation function for the daily log-returns of the closing values of the indices: Dow Jones (DJI) observed from Jan 29, 1985 to Dec 02, 2020; Ibovespa (BVSP) observed from July 24, 1993 to Dec 02, 2020; Nikkei (N225) observed from Jan 05, 1980 to Dec 02, 2020; S\&P 500 (GSPC) observed from Jan 02, 1985 to Dec 02, 2020; Nasdaq (IXIC) observed from Oct 01, 1985 to Dec 02, 2020. We analyze each index for roughly $7 \times 10^3$ days. We note that the returns for each investigated index display an expressive correlation which decays exponentially in time.

\begin{figure}[h]
  \centering
 \includegraphics[scale=0.07]{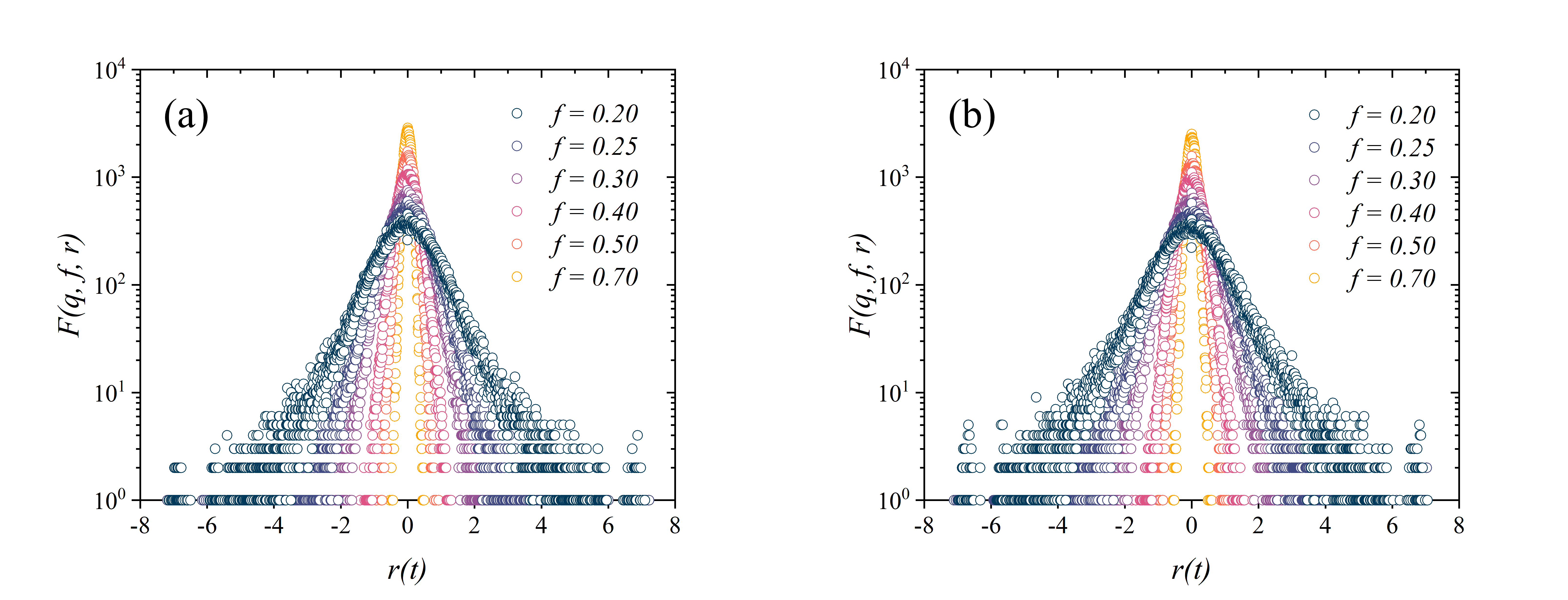}
  \caption{Distribution of logarithmic returns for $10^5$ MCS and several values of the fraction of contrarians $f$: (a) $\left<k\right> = 6$ and $q = 0.240$, and (b) $\left<k\right> = 8$ and $q = 0.275$.}
  \label{Histograms}
\end{figure}

Figure \ref{Histograms} displays the histogram of the log-returns for two sets of parameters and several values of the noise contrarian fraction $f$ in $10^5$ MCS. The real-world market systems display fat-tailed distributions as a reflection of lower non-zero probabilities of obtaining above or below-average returns. We observe such behavior only for lower values of the fraction of contrarians $f$, substantiating a real-world market region for our agent-based dynamics. To quantify the return distributions, we perform a statistical analysis of the data. We obtain that the kurtosis $\textrm{K}(\left<k\right>, f)$ for $f = 0.20$ (strong market phase) are $\textrm{K}(6, 0.20) = 5.85$ and $\textrm{K}(8, 0.20) = 4.98$. For $f = 0.70$ (weak market phase), $\textrm{K}(6, 0.70) = 2.41$ and $\textrm{K}(8, 0.70) = 2.37$. We remark that for a normal distribution $\textrm{K} = 3$, thus, increasing the fraction of contrarians $f$ gradually shifts the system's behavior, from a leptokurtic (fat-tailed) regime, to a mesokurtic (Gaussian) regime.

\begin{figure}[htpb]
  \centering
\includegraphics[scale=0.07]{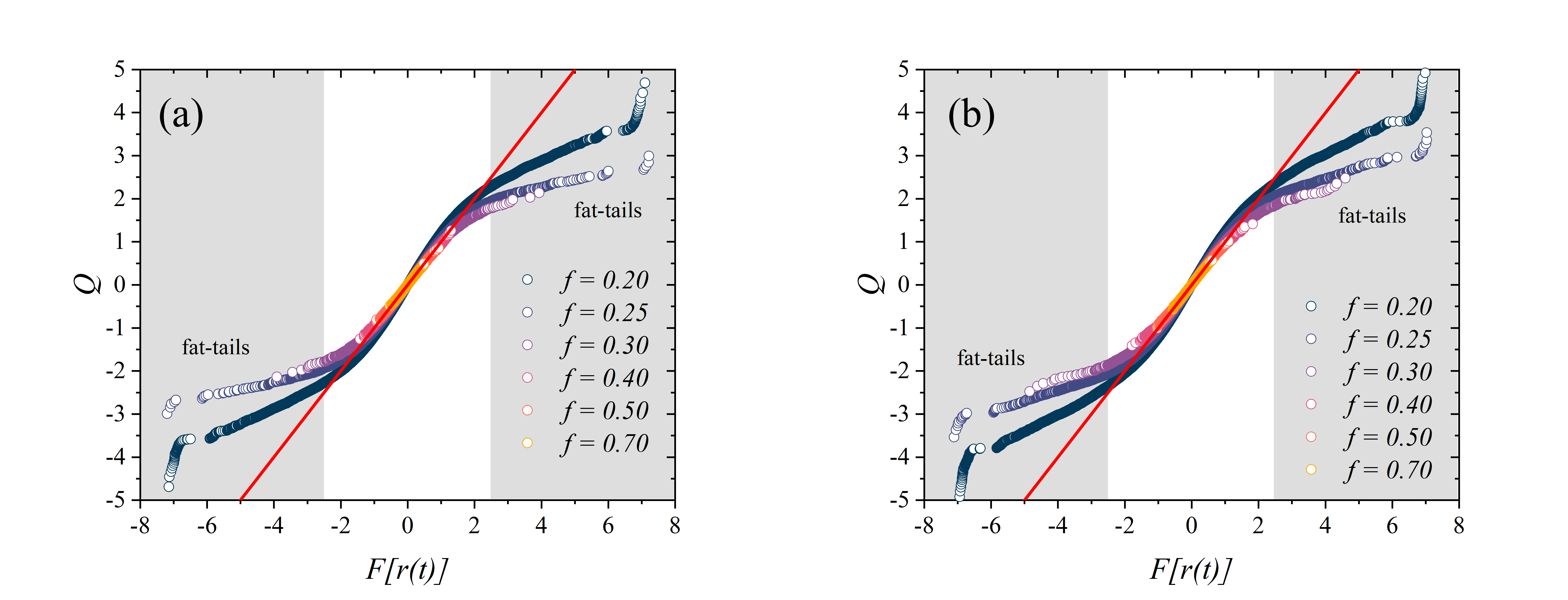}
  \caption{Normal quantile-quantile plots of the logarithmic return distributions for (a) $\left<k\right> = 6$ and $q = 0.240$, and (b) $\left<k\right> = 8$ and $q = 0.275$. Here we use $10^5$ MCS.}
  \label{QQ-Plots}
\end{figure}

To qualify the distributions of Figure \ref{Histograms}, we perform a comparative normal quantile-quantile (Q-Q) plot. Fig. \ref{QQ-Plots} displays Q-Q plots for the distribution of log-returns using several values for the fraction of fundamentalists $f$. The red line represents the expected results of a Gaussian distribution, and if a particular distribution exhibits similar behavior, its data points should lie on that reference line. For lower values of $f$, for instance $f \leq 0.40$, both Figures \ref{QQ-Plots}(a) and \ref{QQ-Plots}(b) display a non-linear behavior in the Q-Q plot, thus indicating that the distributions of log-returns feature fat-tails, and therefore, non-Gaussian. In such distributions, one observes a greater chance of obtaining high return values than the expected for normal distributions, consistent with real-world market systems \cite{stanley2000introduction}.

\begin{figure}[htpb]
  \centering
\includegraphics[scale=0.07]{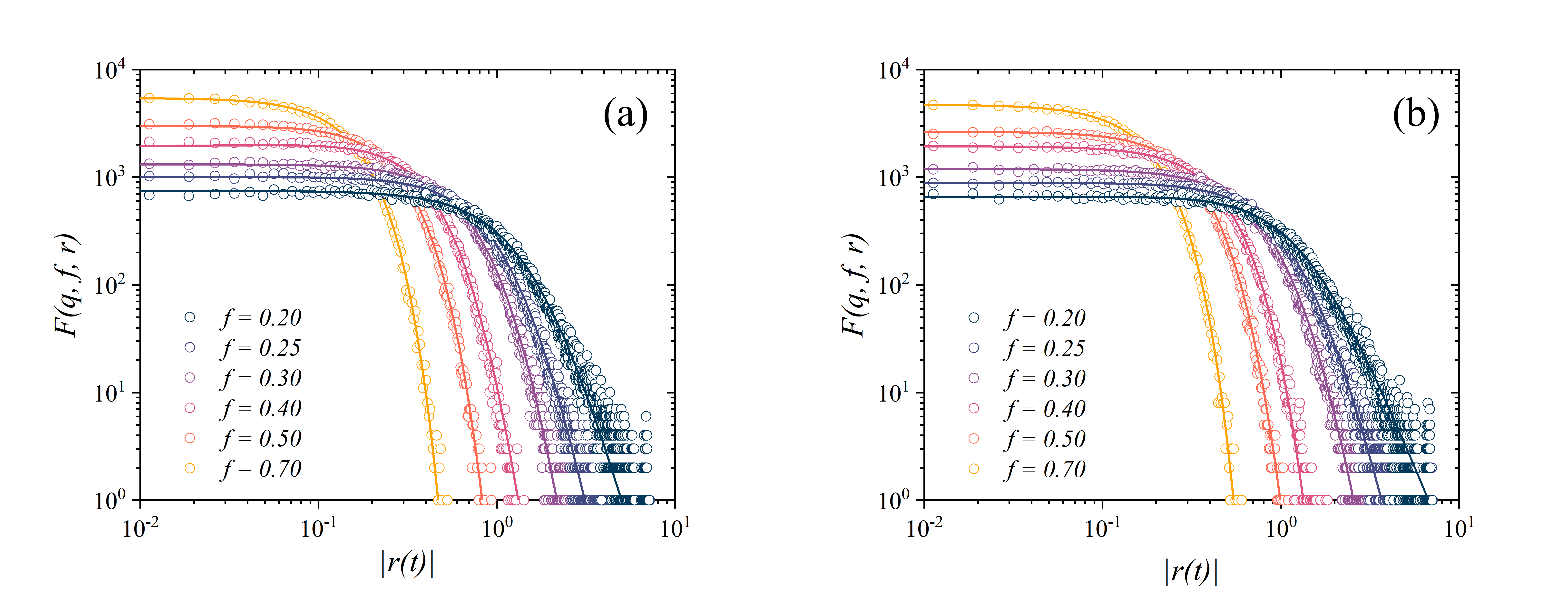}
  \caption{Plot of the absolute log-return distributions $F(q, \ f, \ r)$ for several values of the fraction $f$ and $10^5$ MCS . The lines for $f \leq 0.40$ correspond to Student's t fits, whereas the lines for $f \geq 0.5$ correspond to Gaussian fits. We use (a) $\left<k\right> = 6$ and $q = 0.240$, and (b) $\left<k\right> = 8$ with $q = 0.275$.}
  \label{Student´s t}
\end{figure}

\begin{figure}[htpb]
  \centering
\includegraphics[scale=0.07]{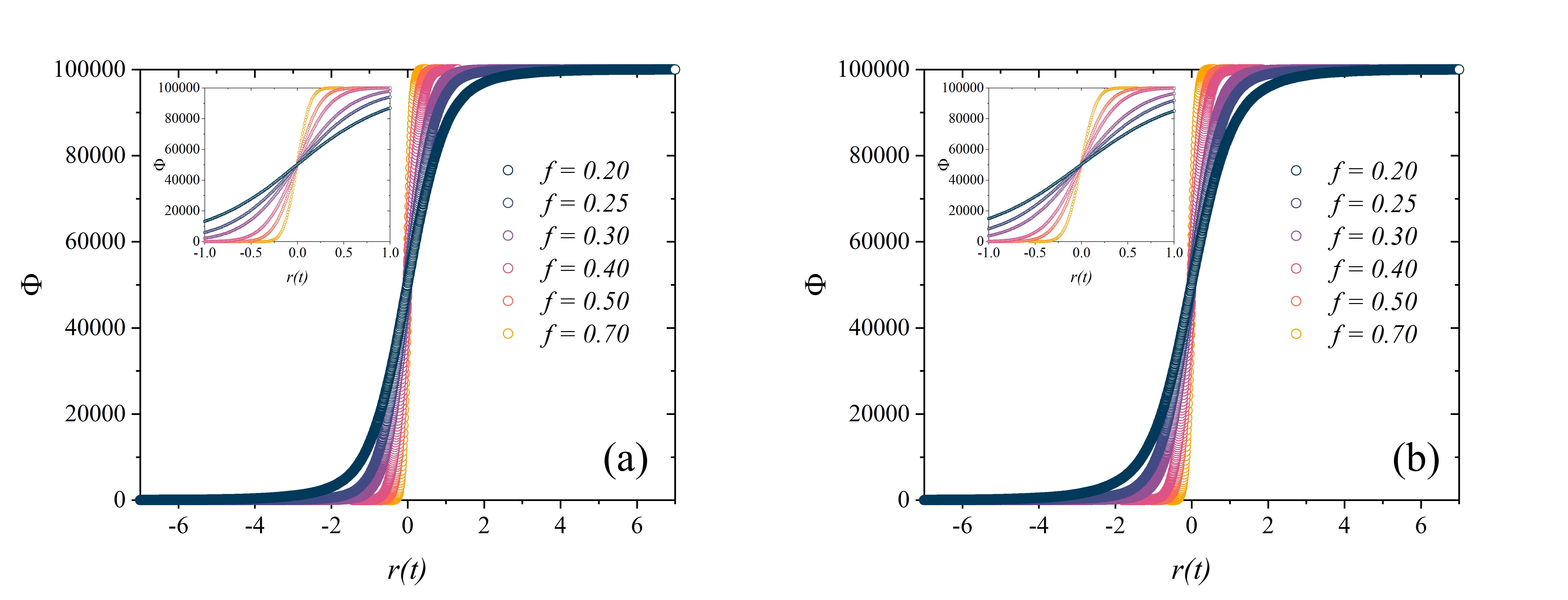}
  \caption{Plot of the cumulative distribution $\Phi$ of log-returns in $10^5$ MCS for (a) $\left<k\right> = 6$ and $q = 0.240$, and (b) $\left<k\right> = 8$ and $q = 0.275$. In the inset we analyze the details of the plot for  values of $r(t)$ near zero.}
  \label{Cumulative Freq}
\end{figure}

Figure \ref{Student´s t} displays the probability distribution of the absolute log-returns for several values of $f$ in $10^5$ MCS with $\left<k\right> = 6$ and $\left<k\right> = 8$. We find that Student's t-distributions can depict the strong market's phase data ($f \leq 0.40$), while the weak market's phase is well fitted by Gaussian distributions \cite{vilela2019majority}. Assuming that the mean values $\mu$ of the distributions are zero, as illustrated by Figure \ref{Cumulative Freq}, we perform generalized Student's t fits for the data $S(r; \nu, \mu, \sigma) = S(r; \nu, \sigma)$, where $\nu$ is the degree of freedom and $\sigma$ represents the scale. The function $S$ is defined as follows

\begin{equation}
   S(r; \nu, \sigma) = \frac{1}{\sqrt{\nu \sigma^2} \textrm{B}\left(\frac{\nu}{2}, \frac{1}{2}\right)} \left(1 + \frac{r^2}{\nu \sigma^2}\right)^{-\frac{\nu + 1}{2}},
\end{equation}
where

\begin{equation}
    \textrm{B}\left(\frac{\nu}{2}, \frac{1}{2}\right) = \int_0^1 t^{\frac{\nu}{2} - 1} (1 - t)^{-\frac{1}{2}} \textrm{dt}, 
\end{equation}
is the Beta function. Values obtained for the degree $\nu$ and scale $\sigma$ are displayed in Table \ref{Degree-scale table}. As the fraction of contrarians increase, we find that the absolute log-return distributions are well fitted by a Gaussian distribution $g(r)$

\begin{equation}
    g(r) = \frac{1}{\sigma \sqrt{2 \pi}} \ \textrm{e}^{-\frac{1}{2} \left(\frac{r - \mu}{\sigma}\right)^2},
\end{equation}
where $\sigma$ is the standard deviation and $\mu$ is the mean, which we shall consider as zero once more. In Table \ref{Std Dev table}, we display the Gaussian fit information.

\begin{table*}[ht]\centering
\ra{1.3}
\setlength{\tabcolsep}{0.25cm}
\begin{tabular}{@{}c c c c c c c@{}}\toprule
& \phantom{abc} & \multicolumn{2}{c}{$\left<k\right> = 6$, $q = 0.240$} & \phantom{abc} & \multicolumn{2}{c}{$\left<k\right> = 8$, $q = 0.275$}\\
\cmidrule{3-4} \cmidrule{6-7}

\ \   $f$  &&      $\nu$      &      $\sigma$     &&      $\nu$      &      $\sigma$         \\ \midrule
\ \ 0.20 && 5.0 $\pm$ 0.3 & 0.80  $\pm$ 0.01  && 3.0 $\pm$ 0.1 & 0.785  $\pm$ 0.008 \ \  \\
\ \ 0.25 && 7.4  $\pm$ 0.5  & 0.558 $\pm$ 0.006 && 6.3  $\pm$ 0.4  & 0.621 $\pm$ 0.006 \ \ \\
\ \ 0.30 && 12 $\pm$ 2  & 0.444 $\pm$ 0.006 && 10 $\pm$ 1  & 0.559 $\pm$ 0.008 \ \ \\
\ \ 0.40 && 13  $\pm$ 2 & 0.252 $\pm$ 0.005 && 38 $\pm$ 2  & 0.319 $\pm$ 0.006 \ \ \\

\bottomrule
\end{tabular}
\caption{Correlation between the scale $\sigma$ and degree $\nu$ as a function of the fraction of contrarians \textit{f}, the average connectivity $\left<k\right>$ and noise $q$.}
\label{Degree-scale table}
\end{table*}

\begin{table*}[hb]\centering
\ra{1.3}
\setlength{\tabcolsep}{0.25cm}
\begin{tabular}{@{}c c c c c@{}}\toprule
& \phantom{abc} & \multicolumn{1}{c}{$\left<k\right> = 6$, $q = 0.240$} & \phantom{abc} & \multicolumn{1}{c}{$\left<k\right> = 8$, $q = 0.275$}\\
\cmidrule{3-3} \cmidrule{5-5}

\ \   $f$  &&      $\sigma$     &&      $\sigma$\\   \midrule
\ \ 0.50 && 0.1916 $\pm$ 0.0008 && 0.2254  $\pm$ 0.0009  \ \  \\
\ \ 0.70 && 0.1108  $\pm$ 0.0004  && 0.1284 $\pm$ 0.0005 \ \ \\

\bottomrule
\end{tabular}
\caption{Dependence of the standard deviation $\sigma$ on the fraction \textit{f}, on the average connectivity $\left<k\right>$ and on noise parameter $q$.}
\label{Std Dev table}
\end{table*}

Figure \ref{Cumulative Freq} displays the cumulative distribution of logarithmic returns $\Phi$ for the data in $10^5$ MCS, from which we imply that the mean for the distributions remains zero for $\left<k\right> = 6$ with $q = 0.240$, and $\left<k\right> = 8$ with $q = 0.275$ for several values of $f$.

\begin{figure}[htpb]
    \centering
    \makebox[0pt]{\includegraphics[scale=0.7]{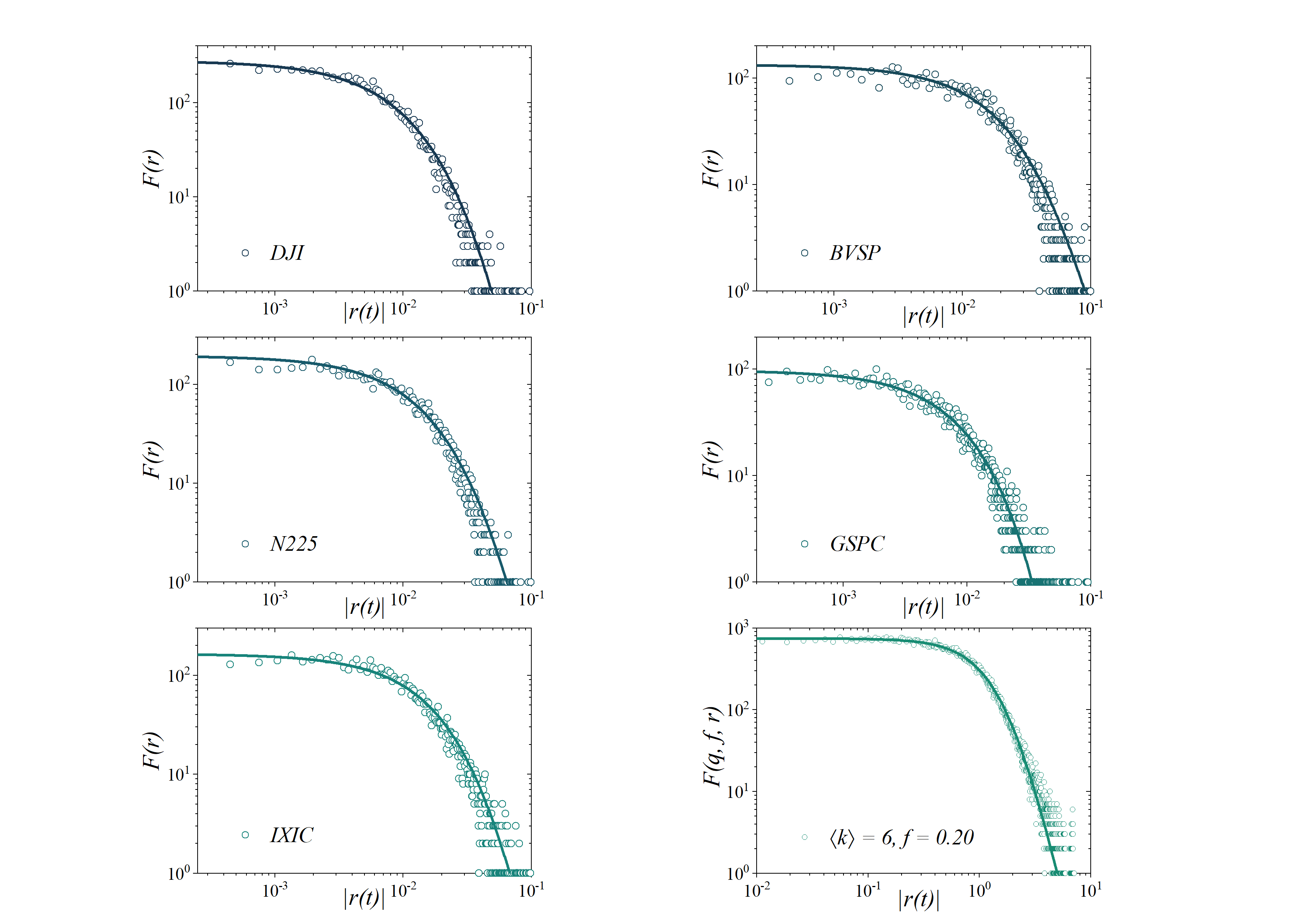}}
    \caption{Plot of the absolute log-return distributions $F(r)$ of the closing values of several financial indices. Also shown is the absolute log-return distributions for $\left<k\right> = 6$, $q = 0.240$, $f = 0.20$. The lines correspond to Student's t fits.}
	\label{StudentFit-RealWorld}
\end{figure}

In Figure \ref{StudentFit-RealWorld}, we show the distribution of the absolute log-returns of the closing values of several financial indices to the results found in our investigation. We analyze each index for roughly $7 \times 10^3$ days and use Student's t-distributions to fit the data, represented by the lines in the plots. In Table \ref{Real World Degree-scale table}, we show the values for the degree $\nu$ and scale $\sigma$. We note that the degree of freedom primarily determines the shape of each distribution, and comparison shows that our model behaves as the expected for real-world financial markets. On the other hand, the scale parameter measures fluctuations around the mean of the distribution, and comparison displays a disparity in the order of magnitude obtained in the fits. Analyzing Figures \ref{fig: SP500 Ret} and \ref{Return}, we observe that the daily log-returns of financial markets are one order of magnitude smaller than the log-returns obtained in our simulations, and thus accountable for the divergence in the results. Hence, we observe that our model represents satisfactorily real-world financial market behavior.

\begin{table*}[ht]\centering
\ra{1.3}
\setlength{\tabcolsep}{0.25cm}
\begin{tabular}{@{}c c c@{}}\toprule

\ \   Index   &      $\nu$       &      $\sigma$    \\ \midrule
\ \ Dow Jones & 9.5 $\pm$ 0.2    & 0.01248  $\pm$ 8 $\times 10^{-5}$ \\
\ \ Ibovespa  & 4.2  $\pm$ 0.3   & 0.0197 $\pm$ 6 $\times 10^{-4}$ \\
\ \ Nikkei    & 15.3 $\pm$ 0.6   & 0.0202 $\pm$ 2 $\times 10^{-4}$ \\
\ \ S\&P 500  & 15.6 $\pm$ 0.5 & 0.0151 $\pm$ 1 $\times 10^{-4}$ \\
\ \ Nasdaq    & 6.5 $\pm$ 0.3    & 0.0180 $\pm$ 3 $\times 10^{-4}$ \\
\bottomrule
\end{tabular}
\caption{Values for the scale $\sigma$ and degree $\nu$ for several financial indices with Student's t fits.}
\label{Real World Degree-scale table}
\end{table*}

\begin{figure}[htb]
    \centering
    \makebox[0pt]{\includegraphics[scale=0.8]{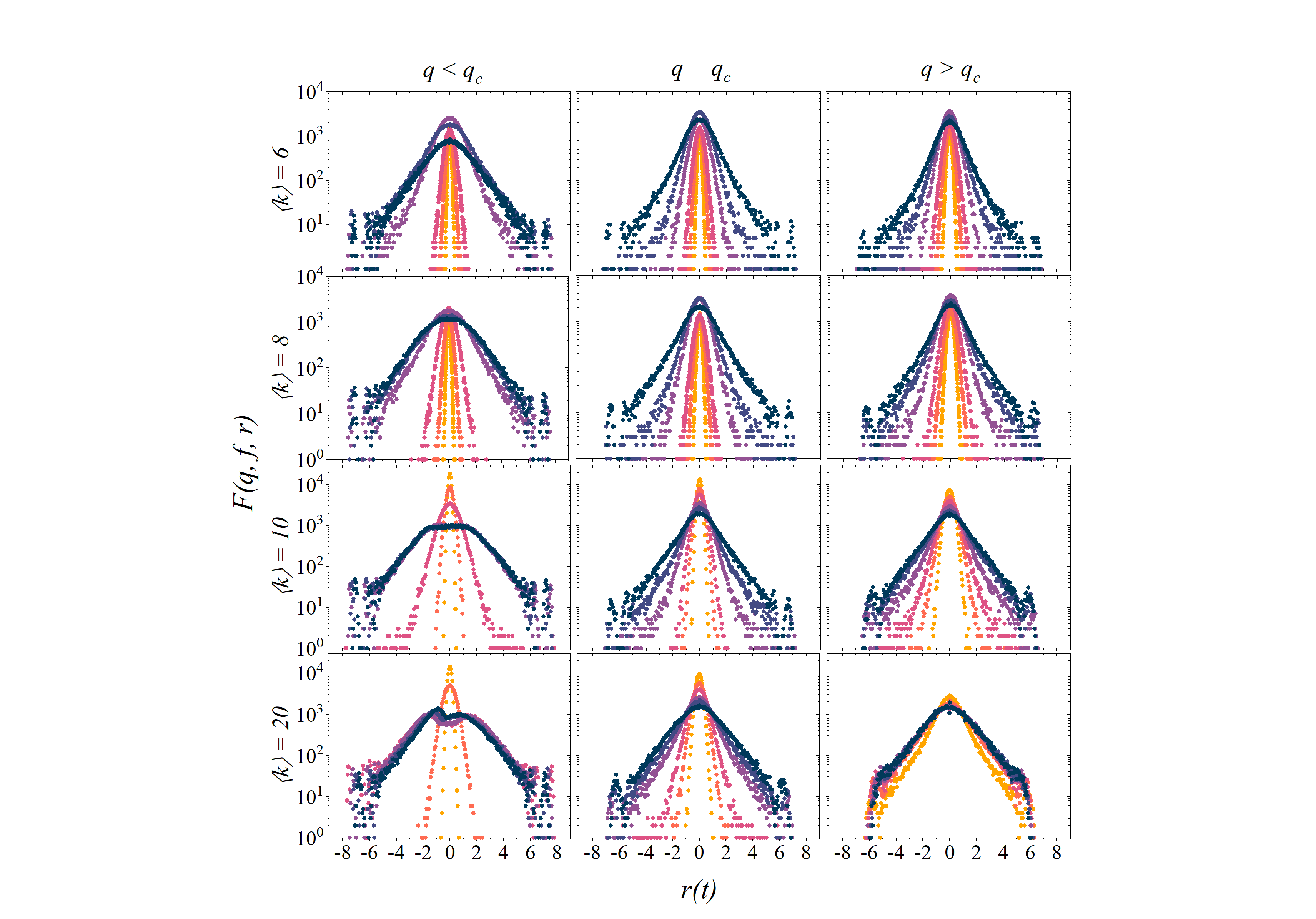}}
    \caption{Distributions of logarithmic returns for several of the average degree of connectivity $\left<k\right>$ in the vicinity of $q_c(\left<k\right>)$ for several values of $f$: 0.20; 0.25; 0.30; 0.50; 0.70 (dark blue, purple, violet, pink, orange, yellow, respectively). Here, we use $10^5$ MCS.}
	\label{Combined Histograms}
\end{figure}

We also perform an extensive investigation of the model for several values of the average connectivity $\left<k\right>$ and different values of the noise parameter $q$. Figure \ref{Combined Histograms} displays a multi-plot table of the histograms of logarithmic returns for several ($q, f,  \left<k\right>$) triplets. Columns correspond to values of $q$, below criticality $q < q_c(\left<k\right>)$ (left), at criticality $q = q_c(\left<k\right>)$ (center) and above criticality with $q > q_c(\left<k\right>)$ (right). In this work, we use the values of $q = q_c(\left<k\right>)$ obtained in previous investigations of the MVM on random graphs, where the fraction of contrarian agents $f$ is set to zero \cite{pereira2005majority}. The values of noise above and below criticality are taken to be $q_c (1 \pm 10\%)$. 

We observe that distributions exhibit fat-tails for lower values of $f$, and especially for small values of $\left<k\right>$, i.e., $\left<k\right> = 6$ and $\left<k\right> = 8$, and at its correspondent critical value of $q$. On the contrary, such behavior is lost for higher values of the average connectivity and values of $q$ that deviate from criticality. This result illustrates and supports our choice for $\left<k\right>$ and $q_c$ used in this investigation.

We conclude that the adoption of random graph networks on the global-vote model for financial markets has demonstrated to be effective. Our model is able to reproduce qualitatively and quantitatively real-world market features for lower values of the average connectivity $\left<k\right>$ and near criticality $q(\left<k\right>) = q_c(\left<k\right>)$. 

We remark that other combination of values for $\left<k\right>$, $f$ and $q$ might yield similar results. Nevertheless, our particular choice adopts simple key ideas: limited and small number of interacting agents, near critical values of the noise parameter $q$ and small number of contrarian agents in the market. Despite the model's simplicity, it has shown its capability of characterizing the mechanisms that drive social behavior and decision making in economic systems.

\section{CONCLUSION AND FINAL REMARKS} \label{conclusion}

This work proposes a generalization of the two-state global-vote model for financial markets on random networks. The global-vote model suggests that any stock market dynamics consist primarily of different agent strategies driven by economic and social interactions. In its standard version, the stock market consists of a heterogeneous population with two distinct kinds of investors: noise traders, who follow the local majority of its neighbors, and noise contrarian traders, influenced by the global minority of the system \cite{vilela2019majority}. 

We aim to investigate the return's distribution dependence on the average connectivity $\left<k\right>$ of the random graph network. We relate variations of the global magnetization of the system to the daily return of a given asset. Our simulations reproduce the typical qualitative and quantitative real-world financial time series. Thus, yielding key features as fat-tailed distributions of returns, volatility clustering and long-term memory volatility. We demonstrate that higher values for the average connectivity $\left<k\right>$, or the noise contrarian fraction $f$, as well as values of $q$ far from criticality, may eliminate similar real-world behavior in our model.

This investigation combines two distinct fields of research - Econophysics and Sociophysics - to better understand the mechanisms at play in the complexity of the human decision-making process, which drives economic systems' dynamics. Consequently, it demonstrates advancements towards comprehending real-world heterogeneous complex systems, such as financial markets.

\newpage
\center{\textbf{Acknowledgements}}

The authors acknowledge financial support from Brazilian and Chinese institutions and funding agents UPE, FACEPE (APQ-0565-1.05/14, APQ-­0707­-1.05/14), CAPES, CNPq, National Natural Science Foundation of China (72071006), the International Postdoctoral Exchange Fellowship Program (20170016) and Beijing Social Science Foundation of China (16JDGLC005). The Boston University Center for Polymer Studies is supported by NSF Grants PHY-1505000, CMMI-1125290,
and CHE-1213217, by DTRA Grant HDTRA1-14-1-0017, and by DOE Contract DE-AC07-05Id14517.

\bibliographystyle{ieeetr}

\newpage 

\end{document}